\documentclass[preprint,aps,prd,showpacs,superscriptaddress,nofootinbib,tightenlines]{revtex4}

\usepackage{amsfonts}
\usepackage{multirow}
\usepackage{mathrsfs}
\usepackage{graphicx}
\usepackage{amsmath}
\usepackage{amssymb}
\usepackage{bm}
\usepackage{bbm}
\usepackage{color}
\usepackage{braket}
\usepackage{slashed}

\def\bfsigma{\mbox{\boldmath $\sigma$}}

\def\bfepsilon{\mbox{\boldmath $\epsilon$}}

\def\OMIT#1{}

\newcommand{\beq}{\begin{equation}}
\newcommand{\eeq}{\end{equation}}
\newcommand{\bqa}{\begin{eqnarray}}
\newcommand{\eqa}{\end{eqnarray}}

\begin{document}
%\preprint{}
%%%%%%%%%%%%%%%%%%%%%%%%%%%%%%%%%%%%%%%%%%%%%%%%%%%%%%%%%%%%%%%%%%%%%%%%%%%%%%
\title{\mbox{}\\[10pt]
Quarkonium decay into photon plus graviton: a golden channel to discriminate General Relativity from Massive Gravity?
}
%%%%%%%%%%%%%%%%%%%%%%%%%%%%%%%%%%%%%%%%%%%%%%%%%%%%%%%%%%%%%%%%%%%%%%%%%%%%

\author{Dong Bai\footnote{dbai@itp.ac.cn}}
\affiliation{School of Physics, Nanjing University, Nanjing, 210092, China\vspace{0.2cm}}

\author{Wen Chen\footnote{chenwen@ihep.ac.cn}}
\affiliation{Institute of High Energy Physics, Chinese Academy of
Sciences, Beijing 100049, China\vspace{0.2cm}}
\affiliation{School of Physics, University of Chinese Academy of Sciences,
Beijing 100049, China\vspace{0.2cm}}

\author{Yu Jia\footnote{jiay@ihep.ac.cn}}
\affiliation{Institute of High Energy Physics, Chinese Academy of Sciences, Beijing 100049, China\vspace{0.2cm}}
\affiliation{School of Physics, University of Chinese Academy of Sciences,
Beijing 100049, China\vspace{0.2cm}}
\affiliation{Center
for High Energy Physics, Peking University, Beijing 100871,
China\vspace{0.2cm}}

\date{\today}
%%%%%%%%%%%%%%%%%%%%%%%%%%%%%%%%%%%%%%%%%%%%%%%%%%%%%%%%%%%%%%%%%%%%%%%%%%%%%%
\begin{abstract}
After the recent historical discovery of gravitational wave,
it is curious to speculate upon the detection prospect of the
quantum graviton in the terrestrial accelerator-based experiment.
We carefully investigate the ``golden" channels, $J/\psi(\Upsilon)\to\gamma+\text{graviton}$,
which can be pursued at \textsf{BESIII} and \textsf{Belle 2} experiments,
by searching for single-photon plus missing energy events.
Within the effective field theory (EFT) framework
of General Relativity (GR) together with Nonrelativistic QCD (NRQCD),
we are capable of making solid predictions for the corresponding decay rates.
It is found that these extremely suppressed decays are completely swamped by the Standard Model
background events $J/\psi (\Upsilon)\to \gamma+\nu\bar{\nu}$.
Meanwhile, we also study these rare decay processes in the context of massive gravity,
and find the respective decay rates in the limit of vanishing graviton mass drastically differ
from their counterparts in GR.
Counterintuitive as the failure of smoothly recovering GR results may look,
our finding is reminiscent of the van Dam-Veltman-Zakharov (vDVZ) discontinuity
widely known in classical gravity, which can be traced to the finite contribution of the
helicity-zero graviton in the massless limit.
Nevertheless, at this stage we are not certain about the fate of the
discontinuity encountered in this work, whether it is merely a pathology or not.
If it could be endowed with some physical significance, the future observation of
these rare decay channels, would, in principle,
shed important light on the nature of gravitation,
whether the graviton is strictly massless, or bears a very small but nonzero mass.
\end{abstract}

%%%%%%%%%%%%%%%%%%%%%%%%%%%%%%%%%%%%%%%%%%%%%%%%%%%%%%%%%%%%%%%%%%%%%%%%%%%%%%

%11.10.Hi Renormalization group evolution of parameters
%12.38.Bx Perturbative calculations
%12.38.Cy Summation of perturbation theory
%12.38.-t Quantum chromodynamics
%12.39.St Factorization
%13.25.Gv hadronic decay of heavy quarkonium
%13.40.Hq Electromagnetic decays
%13.60.Le Meson production
%13.60.Hb Total and inclusive cross sections (including deep-inelastic processes)
%13.87.Fh Fragmentation into hadrons
%04.60.Bc Phenomenology of quantum gravity
%04.50.Kd Modified theories of gravity
%14.40.Pq Heavy quarkonia
%14.70.Kv Gravitons (see also 04.60.-m Quantum gravity)

%%%%%%%%%%%%%%%%%%%%%%%%%%%%%%%%%%%%%%%%%%%%%%%%%%%%%%%%%%%%%%%%%%%%%%%%%%%%%%

\maketitle

\noindent{\color{blue}1. \it Introduction.}
One century after Einstein's avantgarde prediction based on his
newly-formulated General Relativity (GR)~\cite{Einstein:1916},
the recent legendary discovery of gravitational wave from binary black hole mergers by \textsf{LIGO} Scientific Collaboration and \textsf{Virgo} Collaboration~\cite{Abbott:2016blz,Abbott:2016nmj},
has marked a major milestone in humankind's advance in fundamental physics.
A natural question then arises: how and when will we experimentally establish
the existence of the {\it graviton}, the quantum of gravitational field and the force carrier of gravitation?
Due to the extreme weakness of gravitational coupling,
many physicists believe that humankind will never be able to detect graviton.

Looking back in history, one sees that the forerunners of quantum electromagnetism
are much luckier. The making of electromagnetic wave by Hertz in 1887~\cite{Hertz:1893},
was only a quarter of a century after Maxwell's ground-breaking prediction
for the existence of electromagnetic wave~\cite{Maxwell:1865}.
Miraculously, in the same year, Hertz also discovered the photoelectric effect~\cite{Hertz:1887}, which,
from the modern viewpoint, was the direct experimental evidence for the existence of photon,
the force carrier of electromagnetism and the quantum of electromagnetic field.

The path for detecting the graviton is doomed to be much, much more twisty than that
for discovering photon. One may appreciate the difficulty by looking at a simple example.
In his authoritative monograph on gravitation~\cite{Weinberg:1972kfs},
Weinberg estimated the rate of the $3d$-state hydrogen atom transitioning into the $1s$ state via
single-graviton emission to be about $2.5\times 10^{-44}\;\text{sec}^{-1}$,
completely overwhelmed by the spontaneous photon emission rates of order $10^9\;\text{sec}^{-1}$.
So there seems absolutely no chance to detect the graviton in atomic physics laboratories.

This paper reports an exploratory study of graviton-hunting in the terrestrial
high-energy collision experiments. Concretely speaking, we make a comprehensive investigation on,
arguably one of the best channels to discover graviton in the accelerator-based experiments,
$J/\psi(\Upsilon)\to\gamma+\mathcal{G}$ (Henceforth we will use ${\cal G}$
to denote graviton). Since the graviton carries quantum number $J^{PC}=2^{++}$,
this decay process, circumventing the Landau-Yang theorem and allowed by $C$-invariance,
is permissable.
The merit of focusing on vector quarkonia decay is that
they can be copiously produced in $e^+e^-$ collision experiments
with quite clean environment, and also bear very narrow width owing to the OZI-suppression.
The experimental signature is also very simple: a single photon with energy exactly half of the quarkonium mass,
plus invisible events.
We note that \textsf{CLEO}~\cite{Insler:2010jw,Balest:1994ch} and \textsf{BaBar}~\cite{delAmoSanchez:2010ac}
have already placed the upper bounds for the decays $J/\psi (\Upsilon(1S)) \to \gamma+{\rm invisble}$,
but their motivation was to search for the hypothetic dark matter particle, light Higgs and axion.
Here we extend their hunting candidates by adding the graviton.
Roughly speaking, about $10^{10}$ $J/\psi$ samples are accumulated at \textsf{BESIII}, and about $10^{9}$ $\Upsilon(nS)$
($n=1,2,3$) samples will be accumulated in the forthcoming \textsf{Belle 2} experiment, therefore
we expect the upper bound on this rare decay process will continue to improve with respect to Refs.~\cite{Balest:1994ch,Insler:2010jw,delAmoSanchez:2010ac}.

To reliably account for this process, we first need couple the Standard Model (SM)
of particle physics, especially QCD, which is responsible for the binding of heavy quarkonium,
with the gravitation theory, in a consistent manner.
The old folklore was that we do not yet have consistent theory of quantum gravity,
since General Relativity is a nonrenormalizable theory.
Nevertheless, with the increasing popularity of the effective field theory (EFT) motif,
the paradigm has gradually shifted to that, as long as treating GR as the low-energy
EFT with UV cutoff around Planck mass, one is then capable of making consistent and controlled predictions,
regardless of our ignorance of the hitherto unknown UV-completed quantum gravity~\cite{Donoghue:1993eb,Donoghue:1994dn}.
The underlying tenet is intimately analogous to the chiral effective field theory
as the powerful low-energy EFT of QCD~\cite{Weinberg:1978kz}.
Among the beautiful applications of this quantum gravity EFT are
the quantum corrections to Newton's gravitational law between two masses~\cite{Donoghue:1993eb,Donoghue:1994dn,Khriplovich:2002bt, BjerrumBohr:2002kt},
quantum correction to the bending of light in the external gravitational field~\cite{Bjerrum-Bohr:2014zsa,Bjerrum-Bohr:2016hpa,Bai:2016ivl}.

Meanwhile, a model-independent description of the rare decay of vector quarkonium,
a color-singlet meson formed by the heavy quark-antiquark pair (with the quark-model
spectroscopic symbol ${}^3S_1$), necessitates a controlled way of tackling the nonperturbative aspects
of strong interaction. Historically, heavy quarkonium decays have played a key role in establishing
the asymptotic freedom of QCD~\cite{Appelquist:1974zd,DeRujula:1974rkb}.
Due to the non-relativistic nature of the heavy quarks inside quarkonium (the characteristic velocity
of heavy quark $v\ll c$), the decay rates were traditionally
expressed as the squared bound-state wave function at the origin,
which is sensitive to the long-distance binding mechanism,
multiplying the short-distance quark-antiquark annihilation decay rates, which can be accessible to
perturbative $\alpha_s$ expansion.
After the advent of the nonrelativistic QCD (NRQCD) EFT~\cite{Caswell:1985ui},
this intuitive factorization picture has been put on a field-theoretical ground,
and one is allowed to systematically conduct a double expansion in $\alpha_s$ and $v/c$~\cite{Bodwin:1994jh}.
To date a vast number of quarkonium decay and production processes have been
fruitfully tackled using this powerful EFT approach~\cite{Brambilla:2010cs}.
Within the EFT framework of GR together with NRQCD, we are capable of making a
solid prediction for the corresponding decay rates, which, not surprisingly, are extremely small.
Unfortunately, these extremely suppressed decays appear to be completely swamped by the SM
background events $J/\psi (\Upsilon)\to \gamma+\nu\bar{\nu}$, which themselves
already correspond to rather rare decay modes.

The standard theory of gravitation, GR, as an inevitable consequence of the nontrivially
self-interacting massless spin-2 gravitons, has passed many stringent experimental tests~\cite{Patrignani:2016xqp}.
Nevertheless, from the observational side, it remains possible that the graviton might be massive~\cite{deRham:2016nuf}.
A special class of modified gravity theories, massive gravity (MG)~\cite{Hinterbichler:2011tt}, 
has the potential to account for the accelerated expansion of Universe by endowing the graviton 
a Hubble-scale mass rather than introducing the dark energy, which makes it phenomenologically appealing~\cite{Deffayet:2001pu}.
To date, astrophysical bound on the graviton mass is $m_{\cal G}< 6\times 10^{-32}$ eV,
many orders of magnitude tighter than that on the photon mass~\cite{Patrignani:2016xqp}.

Perhaps the most interesting outcome of this work arises from our comparative study
of the rare decays $J/\psi(\Upsilon)\to\gamma+\mathcal{G}$ from both GR and MG.
There we come across a striking finding: the decay rates in the limit of vanishing graviton mass drastically differ
from their counterparts in GR! This discontinuity can be traced to the contribution from the
scalar-polarized graviton, which is absent in GR.
It is natural to interpret this discontinuity as a quantum realization of the van Dam-Veltman-Zakharov (vDVZ) discontinuity~\cite{vanDam:1970vg,Zakharov:1970cc}, which is widely known in classical gravity.
As elucidated by Vainshtein long ago, the vDVZ discontinuity was 
just an artifact of linearized gravity, which can be dissolved by including nonlinearities 
nonperturbatively~\cite{Vainshtein:1972sx}.
Nevertheless, in our case, we are unable of envisaging 
the relevant mechanism to eliminate this discontinuity
even if it is merely a pathology.
On the other hand, this discontinuity may bear some physical significance.
Consequently, if these rare decays were observed one day in distant future (in extremely unlikely situation),
one would then be capable of pinpointing, in principle,
whether the graviton is strictly massless or it has an very small yet nonzero mass.
This criterion, if true, appears to be of fundamental impact on the theory of gravitation.
\\

\noindent{\color{blue}2. \it GR+SM as an EFT for quantum gravity.}
The SM of particle physics and GR for gravity have long been two eminent pillars in fundamental physics.
In modern EFT paradigm, SM and GR are not only not incompatible, but can be fruitfully combined
into a predictive framework for quantum gravity, provided that the probed energy scale
is far below the Planck mass.
The action for the quantum gravity EFT can be built upon the principle of general coordinate invariance,
and can be divided into gravity part and matter part:
%-------------------
\beq
%-------------------
S= S_{\rm grav} + S_{\rm matt} = \int d^4 x \sqrt{-g}({\cal L}_{\rm grav} + {\cal L}_{\rm SM}).
\label{Action:Quantum:Gravity}
%-------------------
\eeq
%-------------------
The pure gravity sector can be organized as the curvature (energy) expansion:
%-------------------
\beq
%-------------------
{\mathcal L}_{\rm grav}= -\Lambda - {2\over \kappa^2} \,R + c_1 R^2 + c_2 R^{\mu\nu} R_{\mu\nu}+\cdots,
\label{LagGR}
%-------------------
\eeq
%-------------------
where $\kappa =\sqrt{32\pi G_N}$, with Newton's constant $G_N=6.709\times 10^{-39}\;\text{GeV}^{-2}$.
$g_{\mu\nu}(x)$ signifies the metric field, $R^{\mu\nu}$ is Ricci tensor, and
$R$ is the Ricci scalar.
The first term corresponds to the cosmological constant, completely immune to
local accelerator experiments.
The effects of quadratic curvature terms are so much suppressed that no meaningful constraints can be
imposed on the dimensionless couplings $c_i$ ($i=1,2$): $c_{1,2} < 10^{74}$~\cite{Stelle:1977ry}.
In this work, suffice it for us to stay with the Einstein-Hilbert action.

The matter action is comprised of all the SM fields that are minimally coupled with gravity:
%-------------------
\beq
%-------------------
{\mathcal L}_{\rm SM} = -\frac{1}{4}g^{\mu\alpha}g^{\nu\beta}F_{\mu\nu} F_{\alpha\beta}-
\frac{1}{4}g^{\mu\alpha}g^{\nu\beta} G^a_{\mu\nu}  G^a_{\alpha\beta}+
\sum_f \bar{q}_f (i\gamma^a e^\mu_a D_\mu - m_f ) q_f+\cdots.
\label{Lagrangian:SM:gravity}
%-------------------
\eeq
%-------------------
For our purpose, we need only retain the matter contents of photon, gluons, and the quarks.
$F_{\mu\nu}$ and $G^a_{\mu\nu}$ represent the field strengths for the photon and gluon,
respectively. The $f$-flavor quark field is denoted by $q_f$, with $m_f$, $e_f$ its mass and electric charge.
$e^\mu_a$ is the vierbein field.
$D_{\mu} =\partial_{\mu} -i e_f e A_{\mu}-ig_s G^a_{\mu}T^a +
\frac{1}{2}\sigma^{ab}\omega_{\mu ab}$ is the covariant derivative acting on the quark fields,
with $\sigma^{ab}={1\over 4} [\gamma^a,\gamma^b]$, and $\omega_{\mu}^ {ab}$ the spin connection.
Apart from the ordinary $SU(3)_{\rm c}\times U(1)_{\rm em}$ gauge couplings,
the last term in $D_{\mu}$ generates the spin-dependent gravitational interaction.

Since we are only interested in terrestrial accelerator experiment, it is legitimate to conduct the
weak-field approximation, by decomposing $g_{\mu\nu}= \eta_{\mu\nu}+\kappa h_{\mu\nu}$,
and treating $h_{\mu\nu}$ as a small spin-2 quantum fluctuation around flat spacetime background.
The most-minus-signature $(+---)$ is adopted for Minkowski metric $\eta_{\mu\nu}$.
Expanding (\ref{Lagrangian:SM:gravity}) to linear order in $h_{\mu\nu}$,
one can schematically express the graviton-matter interactions as
%-------------------
\bqa
%-------------------
&& \mathcal{L}_\text{int} = -{\kappa\over 2} h_{\mu\nu} T^{\mu\nu} =
\mathcal{L}_{\bar{f} f {\cal G}}+\mathcal{L}_{\bar{f} f g {\cal G}}+\mathcal{L}_{\bar{f} f \gamma {\cal G}}+
\mathcal{L}_{gg{\cal G}}+\mathcal{L}_{\gamma\gamma{\cal G}}+\cdots,
%-------------------
\label{linearized:gravity:interaction:lagran}
%-------------------
\eqa
%-------------------
where $T^{\mu\nu}$ is the Belinfante energy-momentum tensor of the SM.
\\

\noindent{\color{blue}3. \it Polarized decay rates and helicity amplitudes.}
Let us choose to work in the $J/\psi$ rest frame. Suppose the spin projection of the $J/\psi$ along the
$\hat{z}$ axis to be $S_z$, the helicities carried by
the outgoing photon and graviton to be $\lambda_1$, $\lambda_2$, respectively.
Let $\theta$ signify the polar angle between the direction of the photon 3-momentum
and the $\hat{z}$ axis. The differential polarized decay rate can be expressed as~\cite{Jacob:1959at,Haber:1994pe}
%-------------------
\beq
%-------------------
{d \Gamma[J/\psi(S_Z)\to \gamma(\lambda_1)+{\cal G}(\lambda_2)]\over d\cos\theta} = {1\over 32\pi M_{J/\psi}}
\left|d^1_{S_z, \lambda_1-\lambda_2}(\theta)\right|^2
\left|{\cal M}_{\lambda_1,\lambda_2}\right|^2,
%-------------------
\label{diff:polar:decay:rate}
%-------------------
\eeq
%-------------------
where ${\cal M}_{\lambda_1,\lambda_2}$ characterizes the helicity amplitude which encodes nonrivial dynamics.
The angular distribution is entailed in the Wigner rotation matrix $d^{j=1}_{m,m^\prime}(\theta)$. Note
the angular momentum conservation constrains that $|\lambda_1-\lambda_2|\leq 1$.

Integrating (\ref{diff:polar:decay:rate}) over the polar angle, and averaging over three $J/\psi$ polarizations,
one finds the integrated decay rate of $J/\psi$ into $\gamma(\lambda_1)+{\cal G}(\lambda_2)$ reads
%-------------------
\beq
%-------------------
 \Gamma[J/\psi \to \gamma(\lambda_1)+{\cal G}(\lambda_2)]  = {1\over 48\pi M_{J/\psi}}
 \left|{\cal M}_{\lambda_1,\lambda_2}\right|^2.
%-------------------
\label{polarized:decay:rate}
%-------------------
\eeq
%-------------------
Since this decay process is mediated by the strong, electromagnetic, gravitational interactions,
the relation ${\cal M}_{-\lambda_1,-\lambda_2}= {\mathcal M}_{\lambda_1,\lambda_2}$ constrained by
parity invariance, can be invoked to reduce the number of independent helicity amplitudes.
\\

\begin{figure}[tb]
\centering
\includegraphics[width=0.8\textwidth]{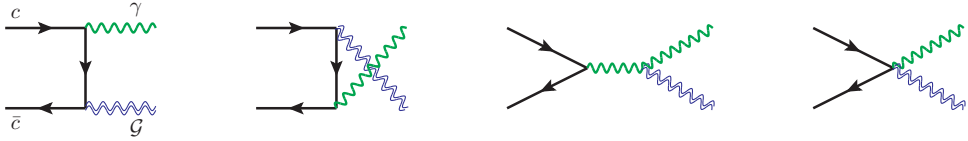}
\caption{Four LO Feynman diagrams for
$c\bar{c}({}^3S^{(1)}_1)\to \gamma+{\cal G}$.}
\label{LO:Feyn:Diag}
\end{figure}

\noindent{\color{blue}4. \it $J/\psi\to \gamma+$massless graviton.}
We are going to apply the NRQCD factorization recipe to the decay $J/\psi\to \gamma+{\cal G}$, starting from SM+GR EFT.
Intuitively, the $c$ and $\bar{c}$ have to get very close to each other in order to annihilate into
a hard photon plus a hard graviton (here {\it hard} means that momentum is of order charm quark mass, $m_c$,
or greater). The computation of this process is in spirit similar to, but more involved than,
the electromagnetic quarkonium decay processes such as $J/\psi\to e^+e^-$ and $\eta_c\to \gamma\gamma$,
with the complication that the graviton couples universally to all the matter fields via the
energy-momentum tensor of SM, including quark, photon and gluon.
The NRQCD short-distance coefficients (SDCs), which encompass {\it hard} quantum fluctuations
emerging in the length scale $\le {1\over m_c}$), can be computed in perturbative QCD owing to
the asymptotic freedom.
The methodology of extracting SDCs is well known, and we refer the interested readers to
Refs.~\cite{Bodwin:1994jh,Petrelli:1997ge,Bodwin:2002hg} for encyclopedic introduction.

In accordance with the Feynman rules generated from the linearized gravitational
interaction encoded in (\ref{linearized:gravity:interaction:lagran}), four quark-level diagrams arise
at the lowest order (LO) in $\alpha_s$, as is shown in Fig.~\ref{LO:Feyn:Diag}.
Let us first consider the static limit, that is, by neglecting the relative momentum
between $c$ and $\bar{c}$. Under dimensional analysis,
the LO helicity amplitudes for $J/\psi\to\gamma+\mathcal{G}$ can be written as
%-------------------
\beq
%-------------------
\mathcal{M}^{\rm LO}_{\pm 1, \pm 2}= e_c e \,\kappa\, {\cal A}^{\rm LO}\,
\sqrt{M_{J/\psi}} \langle 0 \vert \chi^\dagger \bfsigma \cdot \bfepsilon^* \psi \vert J/\psi(\bfepsilon)\rangle,
%-------------------
\label{helicity:ampl:LO:GR}
%-------------------
\eeq
%-------------------
where ${\cal A}^{\rm LO}$ in (\ref{helicity:ampl:LO:GR}) is expected to be a dimensionless ${\cal O}(1)$ number.
Here $\langle 0\vert \chi^\dagger \bfsigma \cdot \bfepsilon^* \psi \vert J/\psi(\bfepsilon)\rangle$
represents the lowest-order $J/\psi$-to-vacuum NRQCD matrix element,
where $\psi,\,\chi$ represent the quark and anti-quark Pauli spinor fields in NRQCD, $\bfepsilon$
represents the polarization vector of $J/\psi$.
This nonperturbative matrix element is often approximated by
$\sqrt{N_c \over 2\pi} R_{J/\psi}(0)$ ($N_c=3$ is the number of colors in QCD),
where $R_{J/\psi}(0)$ denotes the radial wave function at
the origin for the $J/\psi$ in the quark potential model, characterizing the probability amplitude for
$c$ and $\bar{c}$ coincide in space.
Without causing confusion, we will use $R_{J/\psi}(0)$ and the NRQCD matrix element interchangeably.

\begin{figure}[tb]
\centering
\includegraphics[width=0.8\textwidth]{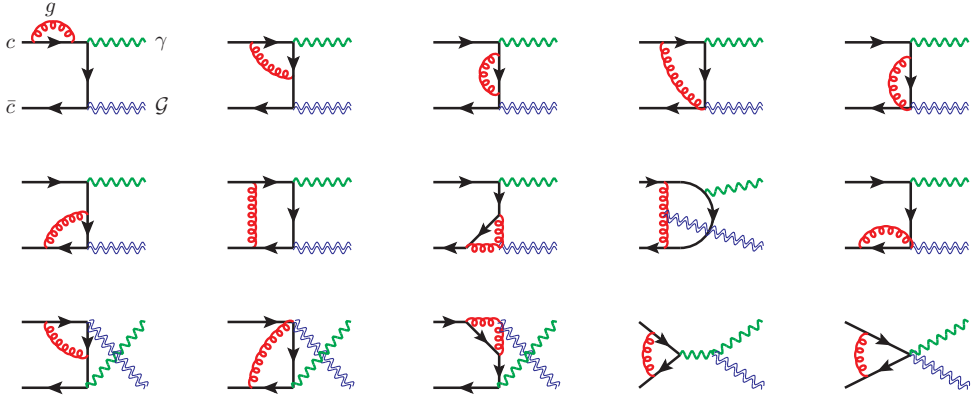}
\caption{Representative Feynman diagrams for
$c\bar{c}({}^3S^{(1)}_1)\to \gamma+{\cal G}$ in NLO in $\alpha_s$.}
\label{NLO:Feyn:Diag}
\end{figure}

As a matter of fact, ${\cal A}^{\rm LO}$ happens to vanish.
This pattern is reminiscent of the exclusive strong decay process $\eta_b\to J/\psi J/\psi$,
where the LO amplitude also vanishes accidently, but becomes nonzero once going to higher order~\cite{Jia:2006rx,Gong:2008ue}.
In a similar vein, we also proceed to the next-to-leading order (NLO) corrections, along the both directions
in $v$ and $\alpha_s$ expansion.
Assuming $\alpha_s\sim v^2$, we ought to include both the order-$\alpha_s$ and order-$v^2$ corrections coherently.
We utilize the all-order-in-$v$ spin projectors developed in \cite{Bodwin:2002hg} to facilitate the
computation of the relativistic correction.
For the calculation of the radiative correction, as is indicated in Fig.~\ref{NLO:Feyn:Diag},
we set the relative momentum between $c$ and $\bar{c}$ to zero prior to conducting loop integration,
which amounts to directly extracting the order-$\alpha_s$ SDC without concerning about Coulomb divergences~\cite{Beneke:1997zp}.
Dimensional regularization is employed to cope with both UV and IR divergences in intermediate steps.
The calculation is also expedited by using \textsf{Mathematica} packages
\textsf{FeynRules}~\cite{Alloul:2013bka}, \textsf{FeynArts}~\cite{Hahn:2000kx}, 
\textsf{FeynCalc}~\cite{Mertig:1990an},  \textsf{\$Apart}~\cite{Feng:2012iq}
and \textsf{FIRE}~\cite{Smirnov:2014hma}.
Ultimately, the NLO helicity amplitudes read
%-------------------
\beq
%-------------------
{\mathcal M}^{\rm NLO}_{\pm 1, \pm 2}=  {e_c e \kappa \over 6\sqrt{\pi}} \sqrt{N_c M_{J/\Psi}}\,R_{J/\psi}(0)
\left(\langle v^2 \rangle_{J/\psi} +
{3 C_F \alpha_s\over 4\pi} (1-4\ln 2)   \right) ,
%-------------------
\label{helicity:ampl:NLO:GR}
%-------------------
\eeq
%-------------------
where the color Casimir $C_F={N_c^2-1\over 2 N_c}$, and the leading relativistic correction is encoded in $\langle v^2 \rangle_{J/\psi}$,
the dimensionless ratio of the following NRQCD matrix elements:
%-------------------
\beq
%-------------------
\langle v^2\rangle_{J/\psi} = {\langle 0 \vert \chi^\dagger \bfsigma \cdot \bfepsilon^*
(-\tfrac{i}{2}\tensor{\mathbf{D}})^{2}\psi \vert J/\psi(\bfepsilon)\rangle\over m_c^2 \langle 0 \vert \chi^\dagger \bfsigma \cdot \bfepsilon^* \psi \vert J/\psi(\bfepsilon)\rangle },
%-------------------
%-------------------
\eeq
%-------------------
where $\mathbf{D}$ denotes the spatial part of the color-covariant derivative.

Some remarks are in order. First, the cancelation of IR divergence in the ${\cal O}(\alpha_s)$
short-distance coefficient in (\ref{helicity:ampl:NLO:GR}), serves a nontrivial
validation of NRQCD approach in a novel setting involving quantum gravity.
Second, both types of NLO corrections are indeed non-vanishing,
yet suffer from severe destructive interference.

Substituting ${\mathcal M}^{\rm NLO}$ in (\ref{helicity:ampl:NLO:GR})
into (\ref{polarized:decay:rate}), taking both helicity configurations
$(\pm 1,\pm 2)$ into account, we then obtain the unpolarized decay rate:
%-------------------
\beq
%-------------------
 \Gamma[J/\psi \to \gamma +{\cal G}]  = {4 e_c^2 \alpha G_N \over 27}   N_c \left|R_{J/\psi}(0)\right|^2
 \left(  \langle v^2 \rangle_{J/\psi} +
{3 C_F \alpha_s\over 4\pi} (1-4\ln 2)  \right)^2.
%-------------------
\label{Jpsi:integrated:decay:rate:GR}
%-------------------
\eeq
%-------------------
It is amusing to see that the coupling constants entering Nature's fundamental interactions,
strong, electroweak, and gravitational, are all packed in a single formula.
\\

{\noindent \color{blue}5. \it $J/\psi\to \gamma+$massive graviton.}
Let us revisit the decay $J/\psi\to \gamma+{\cal G}$
in a special class of infraredly-modified gravitation theories, massive gravity.
For the gravity sector, suffice it for us to add the
Fierz-Pauli mass term to Einstein-Hilbert action:
%-------------------
\beq
%-------------------
\mathcal{L}_\text{\rm grav}= \sqrt{-g}\left[-{2 \over \kappa^2} R+ {m_{\cal G}^2\over 2}
\left(h^{\mu\nu} h_{\mu\nu}-h^\mu_\mu h^\nu_\nu\right)\right],
%-------------------
\label{LagMG}
%-------------------
\eeq
%-------------------
where $m_{\cal G}$ signifies the graviton mass.
For simplicity, we assume the matter fields coupled with the gravity in the same manner as in (\ref{Lagrangian:SM:gravity}).
Since we are only concerned with linearized gravity, we do not bother to specify the self-interactions for gravitons,
so the detailed textures of the influential massive gravity models,
exemplified by the $\Lambda_5$ model~\cite{ArkaniHamed:2002sp} and the ghost-free
de Rham-Gabadadze-Tolley (dRGT) model (also referred to as $\Lambda_3$ model)~\cite{deRham:2010ik,deRham:2010kj},
are largely irrelevant.

Summing (\ref{polarized:decay:rate}) over three independent helicity configurations,
the decay rate can be expressed as
%-------------------
\beq
%-------------------
 \Gamma[J/\psi \to \gamma+{\mathcal G}]  =  {1\over 48\pi M_{J/\psi}}
 \left( 2 \left|{\cal M}_{1,0}\right|^2 + 2 \left|{\cal M}_{1,1}\right|^2 + 2 \left|{\cal M}_{1,2}\right|^2 \right),
%-------------------
\eeq
%-------------------
where we have neglected the tiny graviton mass in phase space integral.

Contrary to the GR case, the amplitudes in massive gravity at LO in $\alpha_s$ and $v$,
no longer vanish. In the $m_{\cal G}\to 0$ limit, each helicity amplitude bears
the following asymptotic behavior:
%-------------------
\begin{subequations}
%-------------------
\bqa
%-------------------
&& {\cal M}^{\rm LO}_{\pm 1, \pm 2} \to   \left({m_{\cal G} \over   M_{J/\psi}}\right)^2
{ e_c e\,\kappa \over 2\sqrt{\pi}} \sqrt{N_c M_{J/\Psi}} R_{J/\psi}(0),
%-------------------
\\
%-------------------
&& {\cal M}^{\rm LO}_{\pm 1, \pm 1} \to  \left({m_{\cal G} \over  M_{J/\psi}}\right)
{e_c e \,\kappa \over 8\sqrt{\pi}} \sqrt{N_c M_{J/\Psi}} R_{J/\psi}(0),
%-------------------
\\
%-------------------
&& {\cal M}^{\rm LO}_{\pm 1,0} \to
{e_c e\,\kappa \over 2 \sqrt{6 \pi}} \sqrt{N_c M_{J/\Psi}} R_{J/\psi}(0).
%-------------------
%-------------------
\eqa
%-------------------
\label{helicity:ampl:massive:gravity}
%-------------------
\end{subequations}
%-------------------
Since $m_{\cal G}/M_{J/\psi} < 2\times 10^{-41}$,
the contributions from the helicity-$\pm 2$ and helicity-$\pm 1$
gravitons are utterly negligible, whereas
the helicity-0 graviton (graviscalar) survives this limit.
As is well known in classical gravity, by employing the St\"{u}kelberg trick,
one readily shows that the coupling of the graviscalar with the trace of the energy-momentum tensor
generally survives the $m_{\cal G}\to 0$ limit~\cite{Hinterbichler:2011tt}.
This is nothing but the very origin of the vDVZ discontinuity.

Therefore, the decay rate is saturated by the graviscalar's contribution
solely:
%-------------------
\beq
%-------------------
 \Gamma[J/\psi \to \gamma+ {\mathcal G}]  =   {2 e_c^2 \alpha G_N \over 9}
 N_c \left|R_{J/\psi}(0)\right|^2.
%-------------------
\label{unpol:decay:rate:massive:gravity}
%-------------------
\eeq
%-------------------
In contrast to the vanishing LO NRQCD amplitude in the massless GR,
here we encounter a nonzero result in the massless limit of the massive gravity.
This symptom can be identified with the quantum counterpart of the
vDVZ discontinuity. We note that
the pattern of discontinuity observed here qualitatively resembles
what is found for the emission of gravitational radiation in classical massive
gravity~\cite{VanNieuwenhuizen:1973qf}.

The vDVZ discontinuity is commonly regarded as an artifact of linearized classical gravity,
which is of no physical significance.
Once the nonlinearities of gravity are taken into account nonperturbatively,
this discontinuity should fade away, so that the principle of continuity
in parameters is recovered~\cite{Vainshtein:1972sx}.
But in our case, it is far from obvious to identify the relevant Vainshtein mechanism to
remove the discontinuity. It is unclear to us how to analytically resum a class of real
emission and loop diagrams incorporating nonlinear multi-graviton interactions,
and ultimately yield a vanishing amplitude as in GR.
Actually, in our case, QCD corrections appear much more important than the quantum
gravitational corrections.

We note that, by treating massive gravity as the quantum EFT, both the
$\Lambda_5$ and $\Lambda_3$ models bear very low UV cutoffs in the particle physics standard,
say, many orders of magnitude smaller than the characteristic hadronic scale of order GeV.
This seems to cast serious doubt on the reliability of our prediction for $J/\psi\to \gamma{\mathcal G}$,
(\ref{unpol:decay:rate:massive:gravity}) in massive gravity,
consequently the discontinuity found in (\ref{unpol:decay:rate:massive:gravity}) might not even
be physically relevant.
Nevertheless, if a weakly-coupled UV completion of massive gravity could be
realized~\cite{deRham:2017xox}, our predictions based on the linearized gravity
would essentially remain intact, since only very rudimentary knowledge about massive graviton
is required in our calculation:
it has five degrees of freedom rather than two.
Frankly speaking, we are not certain about the ultimate fate of this discontinuity.
\\

{\noindent \color{blue}6. \it Numerical predictions.}
To make concrete predictions, we specify the input parameters as follows~\cite{Patrignani:2016xqp,Bodwin:2007fz}:
$e_c={2\over 3}$, $\alpha= 1/137$, $\alpha_s(M_{J/\psi}/2)= 0.30$,
$\Gamma_{J/\Psi}=92.9\;{\rm keV}$,
$\left|R_{J/\psi}(0)\right|^2 = 0.922\;{\rm GeV}^3$,
$\langle v^2\rangle_{J/\psi}= 0.225$.
Substituting these numbers into (\ref{Jpsi:integrated:decay:rate:GR}) and (\ref{unpol:decay:rate:massive:gravity}),
we then predict
%-------------------
\begin{subequations}
%-------------------
\bqa
%-------------------
&& \text{Br}(J/\psi\to\gamma+\mathcal{G})=(2 \sim 8) \times 10^{-40},\qquad\qquad{\rm GR}
%-------------------
\\
%-------------------
&& \text{Br}(J/\psi\to\gamma+\mathcal{G})= 1.4 \times 10^{-37}. \qquad\qquad\qquad{\rm MG}
%-------------------
\eqa
%-------------------
\label{Jpsi:decay:numerics}
%-------------------
\end{subequations}
%-------------------
For the GR prediction, we have estimated the uncertainty by varying the renormalization scale in $\alpha_s$ with from 1 GeV to $M_{J/\psi}$. The relatively large error can be attributed to the
delicate destructive interference between the order-$\alpha_s$ and order-$v^2$ corrections of comparable size.
It is striking to note that the branching fraction in massive gravity is more than
two order-of-magnitude greater than that in GR!

For completeness, we also make the predictions for $\Upsilon(1S)\to \gamma+{\cal G}$.
Substituting $e_b=-{1\over 3}$, taking $\alpha_s(M_{\Upsilon}/2)=0.21$,
and $\Gamma_{\Upsilon(1S)}=54.02\;\text{keV}$,
$\left|R_{\Upsilon(1S}(0)
 \right|^2 = 6.43\;{\rm GeV}^3$,  $\langle v^2\rangle_{\Upsilon}= -0.009$~\cite{Chung:2010vz},
 we find
%-------------------
\begin{subequations}
%-------------------
\bqa
%-------------------
%-------------------
&& \text{Br}(\Upsilon(1S)\to\gamma+\mathcal{G})=(3 \sim 4) \times 10^{-39},\qquad\qquad{\rm GR}
%-------------------
\\
%-------------------
&& \text{Br}(\Upsilon(1S) \to\gamma+\mathcal{G})= 4.1 \times 10^{-37}. \qquad\qquad\qquad{\rm MG}
%-------------------
\eqa
%-------------------
\label{Upsilon:decay:numerics}
%-------------------
\end{subequations}
%-------------------
For the GR prediction, we have estimated the error by varying the renormalization scale
from $1.5$ GeV to $M_{\Upsilon}$. The relatively small uncertainty is due to the much smaller
relativistic correction compared with the radiative correction,
so that the destructive interference does not play a significant role.

Needless to say, both GR and MG predictions in (\ref{Jpsi:decay:numerics}) and (\ref{Upsilon:decay:numerics})
are many orders of magnitude beyond the maximal sensitivity of the
\textsf{BESIII} and \textsf{Belle 2} experiments.

Meanwhile, the really striking message is the dramatic difference between the decay rates
predicted by GR and MG, which may in principle offer an experimental criterion to judge whether graviton
bears a nonzero mass.
\\

{\noindent \color{blue}7. \it Primary SM background.}
One must be concerned with the nuisance that $J/\psi$ non-gravitational rare decays
may dominate over our desired $\gamma+$graviton signals.
Aside from the hypothetical Beyond-SM scenarios that entail dark matter, light Higgs boson and axion,
the major SM background is from $J/\psi\to\gamma+(Z^*\to ) \nu\bar{\nu}$,
where neutrinos simply escape the detector and manifest themselves as missing energy.
$C$-parity invariance dictates that only the axial-vector part of $Zc\bar{c}$ coupling contribute.
At LO in $\alpha_s$ and $v$, the photon energy spectrum is predicted to be~\cite{Yeghiyan:2009xc,Gao:2014yga}:
%-------------------
\bqa
%-------------------
&& {d\Gamma[J/\psi\to \gamma(E_\gamma)+\nu\bar{\nu}]\over d E_\gamma} =
N_\nu {e_c^2 \alpha G_F^2\over 9\pi^3} N_c \left|R_{J/\psi}(0)\right|^2
E_\gamma \left( 1-{E_\gamma \over M_{J/\psi}}\right),
%-------------------
\label{photon:spectrum:Jpsi:neutrino:pairs}
%-------------------
\eqa
%-------------------
where $N_\nu=3$ counts the number of neutrino flavors, $G_F = 1.166\times 10^{-5}~{\rm GeV}^{-2}$
is the Fermi coupling constant. The $Z^0$ exchange has been mimicked by a contact interaction
since $M_{J/\psi}\ll M_{Z}$.
The photon spectrum is a monotonically increasing function with the photon energy $E_\gamma$.
The integrated decay rate $\Gamma[J/\psi\to \gamma \nu\bar{\nu}] =
N_\nu {2\over 27} e_c^2 \alpha G_F^2 M_{J/\psi}^2 N_c \left|R_{J/\psi}(0)\right|^2$, and the corresponding
branching fraction is about $ 10^{-10}$.

Realistic electromagnetic calorimeters always have limited resolution,
typically 2\% of the photon energy at \textsf{BESIII} and \textsf{Belle}.
We can make a rough estimate for these SM background events
that can fake the desired $\gamma+{\cal G}$ signals.
Multiplying (\ref{photon:spectrum:Jpsi:neutrino:pairs})
at $E_{\gamma\:\rm Max}={M_{J/\psi}\over 2}$ by an energy interval $\Delta E_\gamma = 0.02 E_{\gamma\:\rm Max}$,
we found these ``fake" signals bear a branching fraction about $3\times 10^{-12}$.
The observation prospect of these SM events at \textsf{BESIII} is already rather
pessimistic, let alone our desired signals.
Moreover, even if \textsf{BESIII} would operate an unlimited period of time,
the intended $\gamma+{\cal G}$ signals, unfortunately, seem to be completely overwhelmed by these SM background events.

One can also use (\ref{photon:spectrum:Jpsi:neutrino:pairs}) to estimate the decay rate for
$\Upsilon(1S)\to \gamma+\nu\bar{\nu}$ by making straightforward substitutions.
One finds the total branching fraction of this decay channel is about $ 3\times 10^{-9}$,
and the ``fake" single photon plus invisible possess a branching fraction about $8\times 10^{-11}$.
It appears unlikely to observe these SM background events in the 
forthcoming \textsf{Belle 2} experiment.
\\

{\noindent \color{blue}8. \it Summary.} After decades of heroic efforts to search
the gravitational wave, establishing the experimental evidence for graviton
is to become the holy grail in fundamental physics.
It is deserved to be ranked among the top formidable tasks in the history of physics.
Conceivably, the humankind's zest for unraveling the existence of graviton will last
for centuries, if not forever, and the hope will never be extinguished.

In this work, we have investigated, arguably one of the cleanest channels to search for the graviton in
terrestrial accelerator-based experiments such as \textsf{BESIII} and \textsf{Belle 2}:
the decay process $J/\psi (\Upsilon)\to\gamma+\mathcal{G}$, with clean signature of a photon with energy exactly
half of the quarkonium plus invisible.
Based on the EFT paradigm of SM+GR matched onto NRQCD, we are able to make a solid prediction for
this rare decay channel. The extremely suppressed branching fraction, even many orders of magnitude
smaller than the dominant SM background events $J/\psi(\Upsilon)\to \gamma+\nu\bar{\nu}$,
render the detection of these decay modes futile in foreseeable future.
Nevertheless, unceasing experimental endeavour in pushing the upper bound is always rewarding.

From the theoretical angle, our exploration on the rare decays $J/\psi (\Upsilon)\to\gamma+\mathcal{G}$ brings forth
some notable novelties. First, it is an amazing fact that all the fundamental forces in Nature are intertwined
in a single process. Second, our study has illuminated the strength of EFT, in particular,
we have verified the internal consistency for the marriage of the EFT of quantum gravity, with
NRQCD, which describes the strong dynamics for heavy quark bound states.

Most importantly, we have come across one interesting discontinuity when studying
$J/\psi(\Upsilon)\to\gamma+\mathcal{G}$ in the context of massive gravity.
When the graviton mass approaches zero, the resulting decay rate drastically differs from what is
predicted by GR. Since this discontinuity solely stems from the helicity-zero graviton,
it is natural to interpret this as the quantum counterpart of the vDVZ discontinuity.
Nevertheless, we are not certain whether the discontinuity discovered here is merely a
pathology or not. If it is just an artifact of linearized quantum gravity,
we have not been able to envisage the concrete Vainshtein mechanism for its dissolution.
If this discontinuity could be affiliated with some physical significance,
the future measurement of these rare decays, in principle,
would help us to discriminate whether graviton mass is mathematically zero
or not. It is definitely worth further investigation to elucidate the fate of
this discontinuity.

\vspace{.2 cm}
\begin{acknowledgments}
Y.~J. thanks Xian Gao for discussions on massive gravity.
%---------------------------------------------------------------
D.~B. would like to thank Center of High Energy Physics at Peking University for the talk invitation,
after which this work was initialized.
%---------------------------------------------------------------
The work of D.~B. is supported by the National Natural Science Foundation of China (Grant No.~11535004, 11375086, 11120101005, 11175085 and 11235001), by the National Major State Basic Research and Development of China, Grant No.~2016YFE0129300, and by the Science and Technology Development Fund of Macau under Grant No.~068/2011/A.
%---------------------------------------------------------------
The work of W.~C. and Y.~J. is supported in part by the National Natural Science Foundation of China under Grants No.~11475188, No.~11261130311, No.~11621131001 (CRC110 by DGF and NSFC), by the IHEP Innovation Grant under contract number Y4545170Y2, and by the State Key Lab for Electronics and Particle Detectors.
%---------------------------------------------------------------
The Feynman diagrams were prepared with the aid of \textsf{JaxoDraw}~\cite{Binosi:2003yf}.
%---------------------------------------------------------------
\end{acknowledgments}

\end{document}